PAPER

# Multi-Parameter Estimation of Prevalence (MPEP): A Bayesian modelling approach to estimate the prevalence of opioid dependence

Andreas Markoulidakis,[1,2,*] Matthew Hickman,[1,3,4] Nicky J Welton,[1] Loukia Meligkotsidou[5] and Hayley E Jones[1]

[1]Population Health Sciences, Bristol Medical School, University of Bristol, Bristol, UK, [2]Public Health Scotland, UK, [3]Glasgow Caledonian University, UK, [4]National Drug and Alcohol Research Centre, University of New South Wales, Australia and [5]Department of Mathematics, University of Athens, Athens, Greece

*Corresponding author. andreas.markoulidakis@bristol.ac.uk

## Abstract

Estimating the number of the number of people from hidden and/or marginalised populations — such as people dependent on opioids or cocaine — is important to guide policy decisions and provision of harm reduction services. Methods such as capture-recapture are widely used, but rely on assumptions that are often violated and not feasible in specific applications. We describe a Bayesian modelling approach called Multi-Parameter Estimation of Prevalence (MPEP). The MPEP approach leverages routinely collected administrative data, starting from a large baseline cohort of individuals from the population of interest and linked events, to estimate the full size of the target population. When multiple event types are included, the approach enables checking of the consistency of evidence about prevalence from different event types. Additional evidence can be incorporated where inconsistencies are identified. In this article, we summarize the general framework of MPEP, with focus on the most recent version, with improved computational efficiency (implemented in STAN). We also explore several extensions to the model that help us understand the sensitivity of the results to modelling assumptions or identify potential sources of bias. We demonstrate the MPEP approach through a case study estimating the prevalence of opioid dependence in Scotland each year from 2014 to 2022.

**Key words:** Bayesian Methods, Population size estimation, Indirect estimation, Problem drug use, Multiplier methods

## 1. Introduction

Estimating the size of specific populations — such as individuals with opioid dependence — is crucial for understanding population characteristics and informing public health strategies aimed at reducing drug-related harms. Traditional population surveys, however, fail to provide reliable estimates for marginalised populations associated with social stigma, leading to under-reporting and therefore under-estimation of prevalence [15, 32]. Indirect estimation methods, such as capture-recapture and multiplier methods, have been proposed as alternative methods to produce population size estimates in this scenario, but come with notable limitations [14, 16, 17].

Capture-recapture methods involve linking multiple datasets in which individuals from the target population appear (e.g., arrests for drug possession or use, other criminal justice records, and treatment e.g. opioid agonist therapy (OAT) records) and modelling the overlap between these datasets to estimate the number of individuals not captured in any of these sources — thereby estimating the total population size. While capture-recapture estimation can be implemented using relatively accessible log-linear models, this approach can yield sensitive and biased estimates if model assumptions are violated, which is often the case if the true dependence structure is complex. Although it is possible to account for some dependencies between data sources (e.g., OAT referrals from the criminal justice system) and variation in 'capture' probabilities, the approach has been shown to be vulnerable to structural misspecification [18, 19]. Furthermore, accommodating the complexity of the true dependence structure can require interaction terms that cannot be estimated (i.e. the model may not be identifiable), further limiting the applicability of the method [17].

Multiplier methods estimate population size by taking a known count of events (e.g., opioid overdoses) — often called the 'benchmark' — and applying a multiplier, the inverse of the expected event rate among the population of interest [14]. Although conceptually simple, this method relies on accurate and unbiased estimates of event rates, which are often drawn from different regions or time periods with little evidence that they are representative/apply to the specific place and time [3]. Additionally, validity of estimates produced using multiplier methods is dependent on an assumption that the event type is specific to the target population, which may not always be the case [37].

This paper describes and demonstrates an alternative Bayesian modelling approach to prevalence estimation, 'Multi-Parameter Estimation of Prevalence' (MPEP). MPEP is a type of Multi-Parameter Evidence Synthesis (MPES), a class of Bayesian models which combine multiple and potentially diverse sources of evidence by linking the likelihood for each data point via shared underlying parameters [1]. MPES models provide a robust framework to integrate different and potentially diverse types of data and make inferences that are consistent with all relevant data.The most widely known form of MPES is network meta-analysis, where direct and indirect evidence are combined to compare the effectiveness and safety of mulitple interventions [5, 24]. MPES applications to estimate the size of communities of interest include estimating the *number of people and prevalence of HIV* [27], *hepatitis C virus* [33, 13, 29], and *Chlamydia trachomatis* [30].

We will use the term MPEP to describe specifically a type of MPES model developed to estimate the prevalence of opioid dependence — or other community of interest — using large linked data (e.g. Opioid Agonist Therapy [OAT] prescriptions, linked to opioid-related hospital admissions and death records). The starting point for MPEP is identification of a large baseline cohort of people from the community of interest, based on administrative data such as OAT prescription records. As is the case with capture-recapture, our goal is then to estimate the size of the extra (unobserved) population, and hence the overall population size or prevalence. Data on the baseline cohort are linked (ideally at least two types of) events that are specific to the population of interest, allowing computation of event rate models among those in the baseline cohort. By identifying events with the same definitions that are not linked to the cohort, and making assumptions about the relationship between event rates among people from the community of interest who were included in versus not in the baseline cohort (e.g. that the event rates are equal within the same demographic group), the size of this extra population is estimated. Using multiple types of events allows for checking consistency of evidence, increasing robustness of findings or facilitating extensions of the model to address any biases that are identified [8, 23]. Simultaneous regression models on event rates and the 'extra' prevalence are fitted, allowing for variation by key factors, while also borrowing strength across similar groups.

While MPEP infers population size from numbers of events such as opioid-related deaths, it has several distinct advantages over multiplier methods. Notably, by simultaneously modelling multiple event types, MPEP allows us to assess internal consistency and improve robustness of estimates. Unlike traditional multipliers, MPEP uses linked administrative data to estimate event rates that are specific to the correct time period and location, and further refines these by estimating rates by each demographic group and treatment status. The approach therefore accommodates heterogeneity in event rates across subgroups. Notably, MPEP estimates only the unobserved portion of the population, generally leading to more precise prevalence estimates compared to traditional multiplier methods. Variants of the MPEP approach have been applied to estimate the prevalence of opioid dependence in several regions, including England [17], New South Wales, Australia [8], Ohio [7], Massachusetts [35], and Scotland [23].

In this article, we describe the components of the MPEP approach within a generic framework, outlining updates since its original conception and previous applications, and focusing on its most recent and efficient form. The version of MPEP described here offers computational efficiencies by fitting a regression model only on the unobserved part of prevalence — that is, the part of the population not observed in the baseline cohort — and implementing the model in STAN, which enables more targeted and efficient sampling from the target distribution, reducing computation time and memory requirements [4]. This change to the model to regress on the unobserved prevalence substantially improved computational efficiency, reducing run times from approximately five days to six hours in JAGS. Transitioning from JAGS to STAN yielded substantial further improvements, with sampling completed in under 15 minutes. Convergence was assessed using the Gelman–Rubin statistic ($\hat{R} < 1.05$) [10] and an effective sample size of at least 400 for all parameters used in the models [34].

Furthermore, demonstrate assessment of the consistency of evidence across different event types[28] within the MPEP, and demonstrate the flexibility and adaptability of the approach to specific settings, exploring various practical model extensions, including:

- adjusting for potential biases in event data,
- accommodating sparse data and overdispersion, and
- adjusting for misclassification of events among observed and unobserved populations.

Throughout the paper, we demonstrate MPEP using an application to estimate the prevalence of opioid dependence in Scotland between 2014 and 2022 [22]. Our primary goal is to illustrate the general modelling approach, rather than to provide a detailed interpretation of the results.

## 2. Multi-Parameter Estimation of Prevalence

### 2.1. General Description of MPEP

Figure 1 is a simplified schematic demonstrating the idea of MPEP. A large baseline cohort of individuals from the target population is identified, with data linked to specific events. To estimate the size of the total population, we need to estimate the number of people from the target population who are outside of the baseline cohort. Herein, we refer to this as the 'extra population'. We avoid the term 'unobserved', because some of these individuals will be known to other services. The size of the extra population is estimated from numbers of events of the same definition that occurred outside of the cohort, via assumptions about the relationship between event rates in the baseline cohort and in the extra population.

Simultaneous regression models are fitted to event rates and to 'extra' prevalence — the people from the community of interest who are not in the baseline cohort divided by the size of general population — allowing for variation in event rates and extra prevalence by factors such as year, treatment status, age group, and sex. The total estimated population size is the sum of the size of the baseline cohort and the inferred unobserved population. A major advantage of the MPEP approach over traditional multipliers is its capacity to account for variability in event rates based on observed covariates — particularly across treatment periods (e.g., 'on' vs. 'off' treatment if feasible). This feature is especially critical when modelling mortality among individuals with opioid dependence which can show 3 fold difference between in and out of treatment [25].

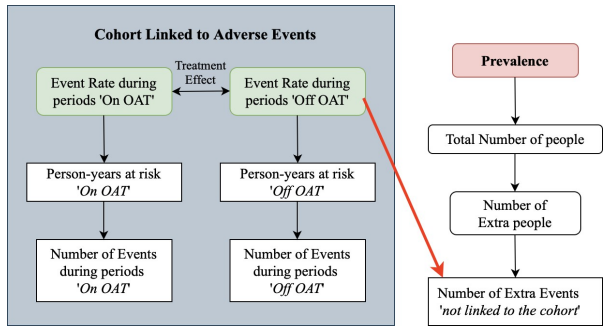

**Fig. 1.** This diagram illustrates the fundamental principle of MPEP: the total population of interest is composed of two groups — the observed baseline cohort and the 'extra' (unobserved) population to be estimated. By making assumptions about the relationship between event rates in the baseline cohort and those in the extra population, the size of the total population is inferred.

Key assumptions of MPEP are:

1. For each demographic group, the rates of events among the unobserved population are assumed to be equivalent to those observed in the baseline cohort during periods when individuals were not receiving OAT.
2. The events modelled are specific to the population of interest — individuals with opioid dependence. That is, the events modeled are assumed to occur exclusively among this population, with no misclassification.
3. The baseline cohort comprises all individuals who have received OAT.
4. All individuals in the baseline cohort are opioid-dependent.

Although these assumptions were tailored for applications of MPEP in estimating the size of opioid-dependent populations, in section 4, we will discuss how the approach can be adapted or extended to relax or replace some of these assumptions.

In the remainder of this section, we describe the key components of the MPEP approach, including defining the baseline cohort; identification of specific event types for modelling; models for event rates within the cohort; model for event not linked to the cohort; model for the rate of exiting the population of interest; specification of the relationship between event rates in versus outside of the baseline cohort; and simultaneous regression modelling of 'extra' prevalence.

## 2.2. Defining the Baseline Cohort

MPEP relies on modelling event rates within a large observed cohort from the population of interest, herein referred to as the baseline cohort. This is a comprehensive list of individuals known to one or more services that interact with the community of interest. In most applications to date [7, 8, 23, 35] the baseline cohort has been defined based on treatment data alone, while in the first application/iteration of the approach [17] the baseline cohort was defined using both treatment and criminal justice data.

Defining the baseline cohort is a key methodological consideration in MPEP studies and requires careful discussion among research teams. Assuming that the cohort is defined based on treatment records, e.g. OAT prescriptions data, an important decision is for how long each individual remains in the target population following their last recorded prescription. For example, in estimating the prevalence of opioid dependence in Scotland, the decision was made to assume that all individuals remained opioid-dependent for five years after their last OAT prescription [23]. More specifically, follow-up began on the day after the last day of an individual's most recent known OAT prescription. Individuals remained in the cohort ('off OAT') for the remainder of that financial year and the subsequent four financial years. Thus, a person was included in the baseline cohort in any year in which they had at least one OAT prescription during that year or within the previous four financial years. In most applications, follow-up time is censored at the earliest of: (1) a pre-specified duration after the last OAT prescription, (2) the date of death (from any cause), or (3) the date of confirmed migration out of the country or region.

We require the baseline cohort to include person-time both 'on' and 'off' treatment, as treatment is expected to influence several event types — for example, opioid-related mortality rates are typically lower while individuals are 'on OAT compared with 'off OAT'. Although defining periods on treatment is generally straightforward (e.g., the duration covered by OAT prescriptions), defining off-treatment time is more complex and requires careful consideration. When specifying the baseline cohort, two issues are especially important:

- How long individuals remain in the cohort after their last recorded prescription, and
- Whether cohort membership should be based solely on treatment records, or whether other data sources indicating drug use or service contact — such as needle-exchange programs — should also be included.

A broader cohort definition may offer advantages, such as increasing off-treatment follow-up time and improving the generalisability of estimated event rates. However, it also restricts the types of events available for estimating the unobserved population. For example, if drug-possession arrests are used to help define cohort membership, they cannot subsequently be used as event types within the MPEP model.

## 2.3. Coding 'On' vs 'Off' treatment

Although classifying off-treatment periods is challenging, defining on-treatment time can be equally difficult. The ideal scenario — having precise prescription start and end dates for every treatment episode — is rarely realised. In the Scotland case study, for example, exact treatment dates were unavailable (as in many similar datasets), and only prescription and reimbursement dates were recorded. Because prescription dates were missing for more than 50% of prescriptions, we adopted a pragmatic approach in which treatment episodes were defined using reimbursement dates, which always correspond to the final day of the month in which a prescription ends. Following the algorithm described by McAuley et al. [25], we defined each treatment episode as beginning 60 days before, and ending 12 days before, the reimbursement date. Consecutive reimbursement dates separated by fewer than 62 days were treated as part of a single continuous treatment episode, with the intervening time classified as on-treatment. Gaps of 62 days or more were classified as off-treatment, and any remaining follow-up time outside these episodes was likewise coded as off-treatment.

Analysts may seek to refine the algorithm used to define treatment episodes — for example, by imputing missing prescription dates and using these to construct more precise treatment timelines. However, this approach introduces additional assumptions about the mechanism of missingness and the factors influencing whether prescription dates are recorded. In our setting, for instance, prescriptions issued outside general practice were systematically missing. Consequently, analysts should prioritise pragmatic methods that align with the structure and quality of their data, drawing on all available sources to improve episode-definition rules and rigorously test their underlying assumptions.

### 2.4. Identification of Specific Event types for Modelling

The baseline cohort data needs to be linked to (ideally at least two types of) event that are specific to the population of interest — e.g. drug-related mortality, non-fatal overdoses admitted to hospital or attended by ambulances, or arrests for drug possession.

Because the size of the unobserved/extra population is estimated from the number of events occurring outside of the baseline cohort, the events included in the model should be completely specific to the population of interest. That is, events of this definition should not occur among individuals outside this community. Otherwise, the model can be expected to overestimate prevalence, unless the model is extended to account for and adjust for this lack of specificity, which ideally would require informative priors based on additional external information.

Although analysts may aim to tailor event definitions to the population of interest, overly broad or non-specific definitions can inflate estimates of the 'extra' population. For example, including suicides within mortality events would increase the number of events attributed to the 'extra' population, thereby inflating the estimated size of that population. In the New South Wales case study, the team initially proposed a definition for hospital admissions among people with opioid dependence, but descriptive analyses revealed substantial inconsistencies in coding practices and data coverage across emergency departments. This prompted a necessary refinement of the definition: certain ICD-10 codes were excluded to ensure that included admissions were clearly opioid-related and reflected a pattern consistent with dependence syndrome.

### 2.5. Notation

Throughout this section, we describe the MPEP model assuming the data are stratified by age, sex, year, region, and, when applicable, treatment status. This is the stratification used in our worked example.

Let $g$ represent the demographic group (defined by age and sex), $y$ denote the year, and $r$ indicate the region. The (observed) size of the general population of group $g$ in financial year $y$ in region $r$ is denoted by $P_{gyr}$.

We denote the size of the 'baseline cohort' — the part of population of interest that is already known — by $n^c_{gyr}$. We wish to estimate the number of 'extra' people ($n^e_{gyr}$) from the population of interest, who are not part of the baseline cohort. The total number of people in the population of interest in each stratification is given by $N_{gyr} = n^c_{gyr} + n^e_{gyr}$.

We introduce an additional index, $s = 0, 1$, to distinguish between periods when individuals are 'off' ($s = 0$) or 'on' ($s = 1$) treatment.

**Table 1.** Table of notations.

| | |
|---|---|
| $g$ | demographic group |
| $y$ | year |
| $r$ | region |
| $s$ | treatment status, $s = \begin{cases} 0 & \text{if } n \text{ 'off' OAT} \\ 1 & \text{if } n \text{ 'on' OAT} \end{cases}$ |
| $i$ | event type |
| $P_{gyr}$ | general population size, group $\{gyr\}$ |
| $n^c_{gyr}$ | size of the baseline cohort, group $\{gyr\}$ |
| $n^e_{gyr}$ | estimated size of the extra popylation, group $\{gyr\}$ |
| $N_{gyr}$ | total (estimated) size of group $\{gyr\}$; $N_{gyr} = n^c_{gyr} + n^e_{gyr}$ |
| $x^c_{igyrs}$ | number of events of type $i$ *within* the baseline cohort group $\{gyrs\}$ |
| $t_{gyrs}$ | follow-up period, group $\{gyrs\}$ |
| $x^e_{igyr}$ | number of events of type $i$ *outside* the baseline cohort group $\{gyr\}$ |
| $t^e_{gyr}$ | estimated follow-up period, group $\{gyr\}$ |
| $t^d_{gyr}$ | person-years at risk among the extra population who died from a death related tot he population of interest, group $\{gyr\}$ |
| $\lambda^c_{igyrs}$ | event-type $i$ rate *within* the baseline cohort group $\{gyrs\}$ |
| $\lambda^o_{gyr}$ | event rate of other cause mortality, *within* the baseline cohort group $\{gyr\}$ |
| $Prev_{gyr}$ | total prevalence |
| $Prev^c_{gyr}$ | prevalence corresponding to the population of the baseline cohort |
| $Prev^e_{gyr}$ | prevalence corresponding to the extra population |

Table 1 summarizes all notations used throughout section 2.

### 2.6. Models for event rates within the cohort

For each group we observe a number of events among those in the baseline cohort. One could use a single type of event (e.g., deaths) or model multiple types of events (e.g., deaths and hospital admissions). We use the term MPEP to describe the approach when at least two types of events are available, allowing for checking for consistency of evidence. Here, we define $x^c_{igyrs}$ as the number of events of type $i$ observed within the corresponding baseline cohort group during the follow-up period ($t_{gyrs}$).

We initially assume Poisson likelihoods for the observed count data:

$$x^c_{igyrs} \sim Poisson\left(\lambda^c_{igyrs} \cdot t_{gyrs}\right),$$

where $\lambda^c_{igyrs}$ are the rate parameters of the Poisson distribution (i.e. the expected rate of occurrences). Relaxations of this assumption are discussed in Section 4.1. We assumed linear model structures for $log(\lambda^c_{igyrs})$ (see Section 3.2).

### 2.7. Model for events not linked to the cohort

Let $x^e_{igyr}$ be the number of events of type $i$ occurring in group $gyr$ that were not linked to the baseline cohort, during follow-up period $t^e_{gyr}$. These events occur among the extra population ($n^e_{gyr}$).

Under the assumption that the baseline cohort includes all individuals receiving treatment, it follows that all of the 'unobserved' population are 'off' treatment. We assume that the event rate for individuals not included in the cohort is equivalent to the event rate for those observed in the cohort during periods 'off' treatment. The likelihood for $x^e_{igyr}$ is then:

$$x^e_{igyr} \sim \text{Poisson}\left(\lambda^c_{igyro} \cdot t^e_{gyr}\right),$$

where $\lambda^c_{igyro}$ were defined before. The unknown quantity time-at-risk for the 'extra' population ($t^e_{gyr}$) is defined in section 2.9.

### 2.8. Model for the rate of exiting the population of interest

Individuals might exit the population of interest, for reasons not directly related to the events modelled, such as migration out of the country or death from any cause not included as an 'event'. Although such exits are accounted for via censoring within the 'observed' population, we wish to estimate the exit rate, to also adjust for this aiming the 'extra' population.

We define as $x^o_{gyr}$ the number of individuals who exited from the cohort due to any cause other than those included in $x^c_{gyro}$. Specifically, this represents deaths or migrations occurring among individuals in the baseline cohort group $gyr$ while they were not receiving treatment. Here, 'o' stands for 'other cause of exit'.

The 'other cause of exit' rates $\lambda^o_{gyr}$ are estimated from the baseline cohort data, during periods 'off' of treatment. This is because we are interested in the event rate within the baseline cohort, among those out of treatment — under the assumption that the baseline cohort includes everyone in receipt of OAT. Therefore, we assume:

$$x^o_{gyr} \sim \text{Poisson}\left(\lambda^o_{gyr} \cdot t_{gyro}\right).$$

### 2.9. The extra population

The unknown quantity $t^e_{gyr}$ represents the person-years at risk for the 'unobserved' part of the population of interest. This value is assumed to be consistent across all event types, allowing it to be jointly estimated from both data sources. In initial development of the approach [8, 17], it was assumed that $t^e_{gyr} = n^e_{gyr}$, implying that each individual in the unobserved population remained at risk for the entire year. This was refined in our application to estimate prevalence in Scotland, to account for the risk of mortality among this population [23].

We assume therefore that

$$t^e_{gyr} = t^d_{gyr} + \left(n^e_{gyr} - t^d_{gyr}\right) \text{RMST}\left(\lambda^o_{gyr}\right),$$

where $\text{RMST}\left(\lambda^o_{gyr}\right)$ is the restricted mean survival time over a one year time period. $t^d_{gyr}$ represents the person-years at risk among the *extra* individuals in group $gyr$ who died from a death related to the population of interest (e.g. opioid-related deaths) — their deaths contributed to $x^e_{gyr}$. This quantity reflects the aggregated proportion of the year prior to death for individuals in this group. By definition, it follows that $t^d_{gyr} \leq x^e_{gyr}$.

Here, $t^e_{gyr}$ is separated into two components: the time at risk for individuals who, while not part of the baseline cohort, are known to have died from an opioid-related cause during the year ($t^d_{gyr}$) — this is because opioid-related deaths are one of the event types we used in the case study of Scotland —, and the time at risk for the remaining unobserved population of interest. Rather than assuming that the latter group remained at risk for the entire year, we account for mortality from other causes or migration using the restricted mean survival time (RMST) term. Specifically, $\text{RMST}\left(\lambda^o_{gyr}\right)$ represents the average proportion of the year that individuals in this group survived, and is calculated as

$$\text{RMST} = \frac{1 - e^{-\lambda^o_{gyr}}}{\lambda^o_{gyr}},$$

under the assumption of a constant event rate during that year, where $\lambda^o_{gyr}$ represents the rate of other cause exit.

### 2.10. Model for the prevalence of opioid dependence

In previous implementations of MPEP, a regression model was fitted to the total prevalence [8, 17, 23]. However, we instead separately modelled the unobserved prevalence, which significantly improved computational stability, enabling faster model execution and facilitating further refinements, such as incorporating greater population stratification (e.g., expanding the number of regions or age groups).

As $n^e_{gyr} = N_{gyr} - n^c_{gyr}$ and $N_{gyr} = \text{Prev}_{gyr} \cdot P_{gyr}$, the model is fully defined once a specification for the prevalence model, $\text{Prev}_{gyr}$, is provided.

Total prevalence ($\text{Prev}_{gyr}$) is separated into two components (observed and unobserved) as follows:

$$\text{Prev}_{gyr} = \text{Prev}^c_{gyr} + \text{Prev}^e_{gyr}.$$

Here, $\text{Prev}^c_{gyr}$ represents the probability parameter of a Binomial distribution fitted to the number of individuals in the cohort ($n^c_{gyr}$), capturing the proportion of the total population ($P_{gyr}$) that is included in the cohort. This is

$$n^c_{gyr} \sim \text{Binomial}(P_{gyr}, \text{Prev}^c_{gyr})$$

A separate regression model is applied to estimate the 'unobserved' prevalence ($\text{Prev}^e_{gyr}$) — see section 3.3.

To estimate prevalence, the inclusion of at least one type of event data is required. However, using multiple event types (at least two) is recommended, as it allows for joint estimation, improving the stability and reliability of the results.

## 3. Case study: Estimating the prevalence of opioid dependence in Scotland

In the application discussed in this work, we use two types of events, namely opioid-related deaths, and non-fatal opioid-related hospital admissions. This case study is used solely to demonstrate some of the features discussed here, rather than commenting on the estimated prevalence of opioid-dependence in Scotland — for this see [22].

The objective is to estimate both the number and prevalence (as a percentage of the general population) of people with opioid dependence in Scotland, stratified by demographic group sex at the time of birth (male/female), age group (15-34, 35-49, 50-64), financial year (2014/15 to 2022/23), and region (seven NHS Boards and 'rest of Scotland').

For this purpose, we used administrative opioid agonist therapy (OAT) PIS record, which is all community prescriptions for OAT records, linked to death records and overdose-related hospital admissions. These data are part of the Substance Use and Health Intelligence Linked Dataset (SHIeLD). We defined the 'baseline cohort' or observed population with opioid dependence as everyone resident in Scotland, age between 15 and 64, who had received OAT in the current year or any of the four preceding years — based on the assumption that people remain opioid dependent for at least that lost after their last OAT.

Individuals were classified as 'on' or 'off' OAT based on the reimbursement dates of OAT prescriptions [25]. This means each individual can contribute time at risk 'on' and 'off' treatment, based on these dates. Follow-up continued for up to five years after the last recorded OAT prescription or until death (from any cause) or migration out of Scotland — whichever occurred first (censoring). The definitions of opioid-related deaths and non-fatal opioid-related overdoses (hospital admissions) used in the model were selected to be highly specific to the population of interest and were based on clinical advice. For example, we only included accidental deaths, where toxicological testing verified that one of heroin, morphine, methadone or buprenorphine, were implicated in death, excluding those who receive opioid for long time for other purposes (e.g. pain). For further details, see [23].

Data from Public Health Scotland (PHS) used for this case study are publicly available as open data[1] from PHS. Due to the small number of events in the complete stratification (age, sex, year, and region), the available data do not include regional stratification to comply with the PHS regulations.

Implementation

To fit the models discussed in this work we used Stan with vague priors for all parameters. We implemented the model in STAN through R statistical software [4]. Example codes of MPEP in Stan are available in a github reporitory[2].

### 3.1. Covariate selection for regression models

The general MPEP approach specification above includes several regression models: event rate models for each type of event, an exit rate model for other causes, and the 'extra' prevalence model. At a minimum, each regression model discussed in this work includes all the main effects — i.e. age group, sex, year, and (where relevant) treatment status. Inclusion of some important interactions could also be pre-specified in the models based on a priori knowledge — for example, past studies of event rates may have identified important interactions that help explain the model. However, it might be unclear which interaction terms should be included in each model; therefore, appropriate covariate selection is recommended.

An approach to choosing which covariates to be included in each regression model is to perform covariate selection for each model, guided by model fit statistics, such as the Deviance Information Criterion (see Appendix A.3) or the Leave One Out Information Criterion. Once the 'best' model is selected for each event rate component, these structures could be held constant while we repeated the procedure for the 'extra' prevalence model.

In our case study, we employed the posterior mean residual deviance and the Deviance Information Criterion (DIC) (see Appendix A.3) to determine which interactions to include in each regression model. Interactions involving the year or region effects were modeled as random effects.

An alternative approach to using information criterion measures to select the covariates included in the regression models is to use regularized priors (e.g., LASSO estimators). This approach was used in an application of the MPEP approach to estimate the prevalence of opioid dependence in Ohio by Doogan et al. [7].

### 3.2. Event Rates Models

Event rates model for different types of events (e.g. deaths and hospital admissions) could have different structures, as different effects might explain the variability of the event rates for each type of event.

As mentioned earlier, the event rate models included all main effects. Additionally, we included a treatment-by-year interaction term, based on previous evidence of such an interaction in drug-related deaths [25].

The final event rate model for hospital admissions used was:

$$\begin{aligned} log(\lambda_{tgyrs}^{c,deaths}) &= \beta_0^h \\ &+ \beta_1^h \cdot treatment_i \\ &+ \beta_2^h \cdot sex_i \\ &+ \beta_3^h \cdot age_i^2 \\ &+ \beta_4^h \cdot age_i^3 \\ &+ \beta_{5:12}^h \cdot year_i^{2015:2022} \\ &+ \beta_{13:19}^h \cdot region_i^{1:7} \\ &+ \beta_{20}^h \cdot sex_i \cdot age_i^2 \\ &+ \beta_{21}^h \cdot sex_i \cdot age_i^3 \\ &+ \beta_{22}^h \cdot sex_i \cdot treatment_i \\ &+ RE^{h,1} \cdot region_i^{1:7} \cdot treatment_i \\ &+ RE^{h,2} \cdot year_i^{2015:2022} \cdot treatment_i \\ &+ RE^{h,3} \cdot year_i^{2015:2022} \cdot region_i^{1:7} \end{aligned}$$

where $\beta_{5:12}^d \cdot year_i^{2015:2022}$ represents the inner product of the vectors $(\beta_5, \beta_6, \beta_7, \beta_8, \beta_9, \beta_{10}, \beta_{11}, \beta_{12})$ and a vector of length eight ($year_i^{2015:2022}$), with each component being a binary indicator for financial years 2015/16 to 2022/23 — 2014/15 is the baseline year. Similarly, for the effect of region, the vector of length seven, $region_i^{1:7}$, is an indicator of NHS Boards (Ayrshire & Arran, Fife, Grampian, Greater Glasgow & Clyde, Lanarkshire, Lothian, Tayside). In our analyses, we used vague prior distributions for all parameters, e.g., Normal distributions with mean zero and large variance for regression coefficients. A random effects (RE) structure was assumed for any interactions with region or year.

---

[1] https://www.opendata.nhs.scot/dataset/estimated-prevalence-of-opioid-dependence-in-scotland/resource/ead97aa5-307d-4d30-a048-3118f2f963fb

[2] https://github.com/andreasmarkoulidakis/mpep.model.git

Following similar notion, the final event rate model for deaths used was:

$$
\begin{aligned}
log(\lambda_{igyrT}^{c,deaths}) =\ & \beta_0^d \\
& + \beta_1^d \cdot treatment_i \\
& + \beta_2^d \cdot sex_i \\
& + \beta_3^d \cdot age_i^2 \\
& + \beta_4^d \cdot age_i^3 \\
& + \beta_{5:12}^d \cdot year_i^{2015:2022} \\
& + \beta_{13:19}^d \cdot region_i^{1:7} \\
& + RE^d \cdot year_i^{2015:2022} \cdot treatment_i
\end{aligned}
$$

The event rate model for the model 'other cause of exit' was:

$$
\begin{aligned}
log(\lambda_{igyr}^o) =\ & \beta_1^o \\
& + \beta_2^o \cdot sex_i \\
& + \beta_3^o \cdot age_i^2 \\
& + \beta_4^o \cdot age_i^3 \\
& + \beta_{5:12}^o \cdot year_i^{2015:2022} \\
& + \beta_{13:19}^o \cdot region_i^{1:7} \\
& + RE^{o,1} \cdot region_i^{1:7} \cdot age_i^2 \\
& + RE^{o,2} \cdot region_i^{1:7} \cdot age_i^3
\end{aligned}
$$

## 3.3. Prevalence Model

Although the structure of the event rate models differs among the different event types, prevalence is estimated jointly from all types of events. The final model used was:

$$
\begin{aligned}
logit(Prev_{igyr}^e) =\ & \gamma_1 \\
& + \gamma_2 \cdot sex_i \\
& + \gamma_3 \cdot age_i^2 \\
& + \gamma_4 \cdot age_i^3 \\
& + \gamma_{5:12} \cdot year_i^{2015:2022} \\
& + \gamma_{13:19} \cdot region_i^{1:7} \\
& + RE^{p,1} \cdot region_i^{1:7} \cdot age_i^2 \\
& + RE^{p,2} \cdot region_i^{1:7} \cdot age_i^3 \\
& + RE^{p,3} \cdot region_i^{1:7} \cdot sex_i \\
& + RE^{p,4} \cdot year_i^{2015:2022} \cdot age_i^2 \\
& + RE^{p,5} \cdot year_i^{2015:2022} \cdot age_i^3
\end{aligned}
$$

We further examined including the observed prevalence as a regressor in the model fitted to the 'extra prevalence', but adding this term did not improve model performance.

## 4. Model Extensions & Adjustments

### 4.1. Exploring and Addressing Overdispersion and Excess Zero Cells

*4.1.1. More zero counts than expected*
In our case-study we fitted Poisson models on the number of events within and outside the cohort. However, a dataset sometimes contains significantly more zero counts than would be expected under a standard Poisson distribution, especially when modelling regions with very small populations. For example, if we aim to extend regional stratification in Scotland for the estimation of the prevalence of opioid dependence to include NHS boards with small total populations, the events included in MPEP may be sparse, resulting in a high number of zero counts.

Zero-inflated models [39, 40] could be used to account for the excess zeros by assuming that the data come from two distinct sources:

1. A zero-inflation component, representing individuals or cases that can only produce zero counts (e.g., structural zeros).
2. A Poisson-distributed component, representing cases where event counts follow a standard Poisson distribution, including some zeros that occur naturally.

A zero-inflated Poisson (ZIP) model [39, 40] could be used to accommodate these two groups, providing a better fit for data with an overabundance of zeros compared to a regular Poisson model, which assumes all observations arise from a single Poisson process.

*4.1.2. Over-dispersion*
The standard Poisson model, assumes that the variance of the number of events in a given time interval is equal to the mean rate of occurrences. However, sometimes the variance exceeds the mean — a phenomenon known as *overdispersion*. If overdispersion occurs in the data, then Poisson regressions can be replaced with negative binomial ones [12]. Wnag et al., have previously used the negative binomial distribution to model the adverse event counts in Massachusetts [35].

In occasions where both over-dispersion and an excessive number of zero counts occur in the same dataset, the Zero-Inflated Negative Binomial model can be used, to account for both. Furthermore, Hurdle Poisson model could be used if the number of zero counts are considered to be less than the expected under a standard Poisson distribution [38].

For the Scotland case study, Figure 2 presents the estimated prevalence of opioid dependence based on three different models for the number of observed events: (1) Poisson, (2) Zero-Inflated Poisson (ZIP), and (3) Negative Binomial (NB). While the prevalence estimates remain largely consistent across all three models, the NB model — and, in some cases, the ZIP model — produces wider credible intervals, reflecting greater uncertainty in the estimates.

*4.1.3. Assessing model fit*
Table 2 reports the residual deviance for four different models — Poisson, ZIP, NB, and ZINB — broken down by sub-models fitted within the MPEP framework, using the Scotland case study as an example. Each of the seven sub-models was fitted to 432 data points (2 genders × 3 age groups × 9 years × 8 regions). Table 3 reports the residual deviance (see Appendix A.3) for each part of the data for the Scotland example, per year of study, and sub-group (e.g. On OAT, Off OAT, Extra) — based on Poisson models only.

It is evident that the Poisson and ZIP models report residual deviances slightly above 432 for the death and other cause of exit (OCE) models, whereas the residual deviance for hospitalisations is typically higher (with the exception of the extra hospitalisations model). Although the model structures for deaths and hospitalisations differ — determined through the model selection process — the models fitted to hospitalisations among the cohort did not achieve residual deviance values as low as those for deaths. Nonetheless, the average per-point contribution remains within acceptable levels (e.g., for hospitalisations during periods on OAT: 542.7/432 = 1.26).

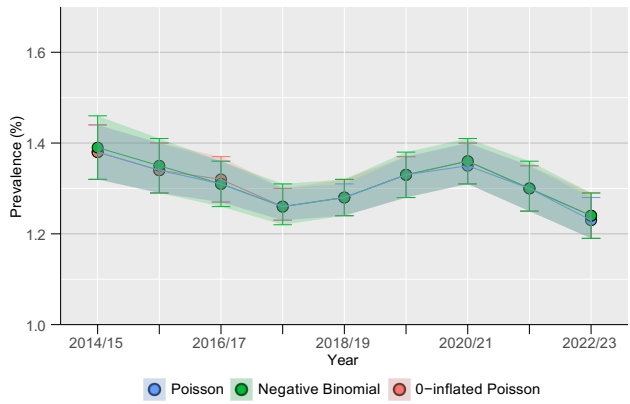

**Fig. 2.** Estimated prevalence (%) of opioid dependence in Scotland; Models: Poisson, Negative Binomial, 0-inflated Poisson (ZIP); 2014/15 to 2022/23

**Table 2.** Breakdown of Residual Deviance; Number of data points per model: 432. Total number of data points: 3024.

|  | Poisson | NB | ZIP | ZINB |
|---|---|---|---|---|
| On OAT: Deaths | 479.6 | 598.1 | 475.7 | 589.5 |
| On OAT: Hospitalisations | 542.7 | 614.6 | 540.5 | 608.8 |
| Off OAT: Deaths | 486.4 | 552 | 483.5 | 545.8 |
| Off OAT: Hospitalisations | 543.2 | 629.8 | 537.1 | 628 |
| OCE | 490.2 | 505.3 | 488 | 502.4 |
| Extra: Deaths | 542 | 671 | 534.7 | 659 |
| Extra: Hospitalisations | 515.8 | 579.2 | 502.9 | 556.6 |
| Total | 3599.9 | 4150.1 | 3562.2 | 4090.1 |

**Table 3.** Breakdown of Residual Deviance; Number of data points per sub-model per year: 48. Number of data points per type of event per year: 144.

|  | On OAT | | Off OAT | | Extra | | Total | |
|---|---|---|---|---|---|---|---|---|
|  | D | H | D | H | D | H | D | H |
| 2014/15 | 37.8 | 67.7 | 54.4 | 64.9 | 44.1 | 49.6 | 136.3 | 182.2 |
| 2015/16 | 60.5 | 52.8 | 42.7 | 50.8 | 47.2 | 44.0 | 150.4 | 147.6 |
| 2016/17 | 51.9 | 59.0 | 55.4 | 54.8 | 75.4 | 58.3 | 182.7 | 172.1 |
| 2017/18 | 57.6 | 65.9 | 45.8 | 70.7 | 57.1 | 50.0 | 160.4 | 186.6 |
| 2018/19 | 63.8 | 69.9 | 54.5 | 69.1 | 55.3 | 61.7 | 173.6 | 200.7 |
| 2019/20 | 53.8 | 53.7 | 45.9 | 69.0 | 76.7 | 71.2 | 176.4 | 193.9 |
| 2020/21 | 51.5 | 58.7 | 58.9 | 41.5 | 69.1 | 78.5 | 179.4 | 178.7 |
| 2021/22 | 51.2 | 62.2 | 72.2 | 69.5 | 57.6 | 44.2 | 181.0 | 175.9 |
| 2022/23 | 51.5 | 52.8 | 56.7 | 52.9 | 59.6 | 58.4 | 167.7 | 164.1 |

Table 4 presents the model fit results for four different models: Poisson, ZIP, NB, and ZINB. Model fit is assessed using the Deviance Information Criterion (DIC), where lower values indicate a better fit — we discuss the DIC further in the appendix. Comparing ZIP vs Poisson, however, overall across the full model there is a 9.4 point reduction in DIC. The ZIP model may therefore be preferred.

On the right hand side of the table, we also report the posterior means of model-specific parameters: the dispersion parameter ($\theta$) for the NB model, the inflation parameter ($\pi$) for the ZIP model, and both ($\theta, \pi$) for the ZINB model. Large values of $\theta$ suggest minimal overdispersion in the data. We see that all of the $\theta$ parameters are large, suggesting little overdispersion, except potentially for the hospital admissions model among those 'off' OAT. Consistently with this, we do not see any overall reduction in DIC from fitting NB relative to Poisson models.

Notably, we did not fit a regression model to the inflation parameter ($\pi$) of the ZIP model. In this application, model selection was used to determine which covariates to include in the Poisson model, and the same regression structure was applied to the event rates of the NB and ZIP models. However, the model could be extended to incorporate covariates also affecting $\pi$. While the primary aim of this work is to introduce these modelling adjustments and highlight the flexibility of MPEP, we recommend conducting a proper model selection study for each model and their parameters to ensure optimal fit and interpretability.

### 4.2. Consistency of Evidence

When multiple types of events are available, jointly estimating prevalence from all sources can enhance the robustness and stability of the estimates. However, it is important to assess the consistency of the estimated prevalence across different event types. Specifically, suppose that two event types, A and B, are used. In that case, it is valuable to determine how similar the prevalence estimates would be if derived solely from event type A versus event type B — that is, to evaluate the degree of agreement between the evidence provided by the two sources.

More formally, we can employ node splitting, a technique first introduced by O'Hagan [11] to identify discrepancies between information sources in Bayesian hierarchical models. Such methods are widely used in network meta-analysis to assess inconsistencies between direct and indirect evidence [5, 24]. We applied a consistency measure (also called 'conflict diagnostic'), proposed by Presanis, et al., 2013 [28], which defines a Bayesian p-value to quantify the similarity between the two estimates. Since, in the application discussed here, we are interested in the total estimated number of people with a given characteristic, we define the continuous quantity $\delta = Prev^e_A - Prev^e_B$ — or equivalently $\delta = N_A - N_B$, where $N_A$ and $N_B$ represent the estimated population size derived from event types A and B, respectively. If the two estimated quantities ($Prev^e_A$ and $Prev^e_B$) are similar, the posterior distribution of $\delta$ should be approximately normal and centered around 0.

We suggest calculating a 'consistency p-value' [28], as:

consistency p-value = $2 \times min\{Pr(\delta > 0), 1 - Pr(\delta > 0)\}$,

which provides a two-tailed test. A value of $c$ close to 0 indicates a low degree of consistency between the posterior distributions of $Prev^e_A$ and $Prev^e_B$, suggesting potential discrepancies in the estimates derived from the two types of events [28].

Using the rule discussed above, one could assess overall consistency for each year, by age, sex, and region, or any specific stratification where analysts suspect inconsistencies.

Figure 3 illustrates the estimated prevalence of opioid dependence in Scotland between the financial years 2014/15 and 2022/23 for individuals aged 15 to 64. The figure presents estimates derived separately from deaths and hospital admissions, as well as a joint estimate incorporating both event types. While estimates from individual data sources are largely

**Table 4.** Breakdown of Deviance Information Criterion. Models: Poisson, Negative Binomial (NB), 0-inflated Poisson (ZIP), 0-inflated Negative Binomial (ZINB).

|  | DIC | | | | Model Parameters | | |
| --- | --- | --- | --- | --- | --- | --- | --- |
|  | Poisson | NB | ZIP | ZINB | NB ($\theta$) | ZIP ($\pi$) | ZINB ($\theta, \pi$) |
| On OAT: Deaths | 1441 | 1444.3 | 1439.4 | 1442.5 | 96.6 | 0.009 | 92.9 , 0.008 |
| On OAT: Hospital Admissions | 1910.7 | 1908.7 | 1908.4 | 1906.7 | 126.2 | 0.007 | 134.8 , 0.006 |
| Off OAT: Deaths | 1630.4 | 1633.6 | 1628.3 | 1631.4 | 453.1 | 0.007 | 474 , 0.007 |
| Off OAT: Hospital Admissions | 1603.5 | 1606.5 | 1602.4 | 1605 | 20.7 | 0.011 | 22.8 , 0.008 |
| OCE | 1964.9 | 1964.9 | 1963.3 | 1963.1 | 177.5 | 0.005 | 196.7 , 0.004 |
| Extra: Deaths | 1744.3 | 1745.2 | 1742.9 | 1743.1 | 60.5 | 0.012 | 71.5 , 0.01 |
| Extra: Hospital Admissions | 1564.1 | 1561.1 | 1564.8 | 1562.8 | 46.4 | 0.018 | 71.6 , 0.015 |
| Total | 11858.9 | 11864.3 | 11849.5 | 11854.6 | | | |

Where $DIC = -2l(y|\theta) + 2pV$, $pV = 2\left[l(y|\theta) - l(y|\bar{\theta})\right]$, $l(y|\theta)$ is the posterior mean of the total log likelihood, and $l(y|\bar{\theta})$ is the log likelihood evaluated at the posterior mean of the parameter $\theta$; OCE = Other Cause Exit; OAT = Opioid Agonist Therapy.

**Table 5.** Estimated prevalence (%) and 95% CrIs of opioid dependence by data source in Scotland; Data Sources: Opioid-related deaths, Opioid-related hospitalisations, full model (both data sources); 2014/15 to 2022/23.

|  | Estimated Prevalence by Data Source | | | |
| --- | --- | --- | --- | --- |
| Year | Deaths | Hospitalisations | Full model (both data sources) | Consistency p-value |
| 2014/15 | 1.44% (1.35%, 1.55%) | 1.35% (1.28%, 1.44%) | 1.38% (1.32%, 1.44%) | 0.20 |
| 2015/16 | 1.37% (1.29%, 1.46%) | 1.34% (1.27%, 1.42%) | 1.34% (1.29%, 1.40%) | 0.68 |
| 2016/17 | 1.32% (1.26%, 1.40%) | 1.34% (1.27%, 1.42%) | 1.31% (1.27%, 1.36%) | 0.72 |
| 2017/18 | 1.31% (1.25%, 1.37%) | 1.25% (1.20%, 1.31%) | 1.26% (1.23%, 1.30%) | 0.20 |
| 2018/19 | 1.30% (1.25%, 1.36%) | 1.27% (1.22%, 1.33%) | 1.28% (1.24%, 1.31%) | 0.50 |
| 2019/20 | 1.37% (1.31%, 1.44%) | 1.31% (1.25%, 1.38%) | 1.33% (1.28%, 1.37%) | 0.30 |
| 2020/21 | 1.40% (1.34%, 1.46%) | 1.34% (1.27%, 1.42%) | 1.35% (1.31%, 1.40%) | 0.40 |
| 2021/22 | 1.34% (1.28%, 1.42%) | 1.26% (1.19%, 1.34%) | 1.30% (1.25%, 1.35%) | 0.20 |
| 2022/23 | 1.29% (1.23%, 1.36%) | 1.17% (1.12%, 1.25%) | 1.23% (1.19%, 1.28%) | 0.04 |

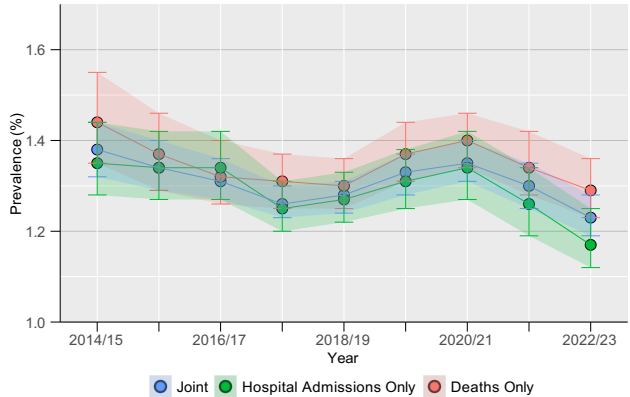

**Fig. 3.** Estimated prevalence (%) of opioid dependence in Scotland; Data Sources: Opioid-related deaths, Opioid-related hospitalisations, Joint (both data sources); 2014/15 to 2022/23

consistent across most years, greater discrepancies are observed in 2022/23.

Table 5 provides the estimated prevalence (expressed as a percentage of the total population) for each data source, with the final column reporting the consistency p-value. Notably, the consistency p-value is close to zero (< 0.05), suggesting significant discrepancies between the estimates obtained from the two event types for 2022/23.

### 4.2.1. Quantification of potential bias in the presence of conflict

Where conflict is detected, it is important to consider potential explanations for it. For example, in the financial year 2022/23 (see table 5), there is a discrepancy between the estimated prevalence based on opioid-related deaths and that based solely on non-fatal opioid-related hospital admissions, with the former being significantly higher. That motivated the exploration of alternative types of events (e.g., ambulance data) in future work, to understand the observed difference.

In some cases, analysts may be able to pinpoint the source of such conflicts and determine whether one data source is known or suspected to be substantially biased. Depending on the circumstances, they may choose to exclude the biased data source or address potential biases by expanding the parameter space and/or incorporating additional data.

The observed inconsistency led us to question the assumption — applied to one of the event types — that the event rate in the unobserved population is equal to the event rate in the observed cohort during periods 'off' OAT. To explore the validity of this assumption, we investigated how much it would need to be violated to ensure consistency between the observed data sources.

We followed the approach of Presanis et al. [26] who discuss strategies for addressing potential conflicts across different sources of evidence. These include the introduction of bias parameters — similar to how we discussed it here —, excluding certain data sources if deemed more appropriate, or extending the model by incorporating additional data sources to inform specific parameters (where possible). In the Scotland case study, we will explore introducing alternative types of events

— such as ambulance attendances that deploy naloxone — to either replace or complement hospital admissions or death records.

In general, let us assume we have two event types, A and B, and we suspect that event type B provides biased evidence of the quantity of interest. A straightforward way to account for this potential bias is to introduce an adjustment term into the event rate model for that event type. This bias term, denoted as $bias_{igyr}$, can be incorporated into the extra event rate definition as follows:

$$\lambda^c_{igyr0} \cdot t^e_{gyr}$$

$$\downarrow$$

$$\lambda^c_{igyr0} \cdot t^e_{gyr} \cdot e^{bias_{igyr}}$$

Importantly, any term allowing for a particular source of bias does not need to be applied universally across all event types ($i$), groups ($g$), years ($y$), and regions ($r$). In addition, similar terms could be incorporated into event rates within the cohort. Instead, it can be selectively introduced only for the specific event type or subset that is suspected to provide biased estimates of the quantity of interest. The bias term can be assigned a non-informative Normal prior with a mean of zero and a large standard deviation, allowing for exploration of potential bias. Alternatively, if prior knowledge or external evidence exists regarding the potential bias, an informative prior can be used to incorporate this information into the model to adjust for bias — this approach is discussed in the next section.

Using the data from the case study conducted in Scotland, we introduce the bias term specifically to adjust the extra event rate for hospital admissions. Figure 4 illustrates the estimated prevalence of opioid dependence in Scotland, comparing estimates derived jointly from deaths and non-fatal hospital admissions, both with and without the bias adjustment, and solely form deaths. The results show that incorporating the bias term brings the estimated prevalence closer to the estimates based solely on deaths, although some differences remain. This indicates that while hospital admissions data still contribute to the prevalence estimation, their influence is moderated by the inclusion of the bias term. Adding bias terms to improve model fit is not an optimal choice; however, it can be useful for identifying where additional information could enhance the model or which sources of evidence might be unreliable.

For the model incorporating the bias term, we applied a non-informative prior, as previously discussed, $Normal(0, 10^2)$. To put this into perspective, the size of the mean of the bias term for the counts corresponding to 2022/23 varied between 0 and 2.7, showing substantial variability. For instance, the 95% Credible Interval (CrI) for one bias term with a mean value of 2.7 ranged from 0.56 to 10.27, highlighting considerable uncertainty. The inclusion of such a term could influence the estimated quantity of interest, in our example, potentially causing the prevalence estimates to be driven almost entirely by death events.

The introduction of bias parameters, as discussed here, is often associated with the need to re-evaluate model assumptions. This process may also benefit from the incorporation of external evidence, either through informative priors or the inclusion of supplementary data sources, where available.

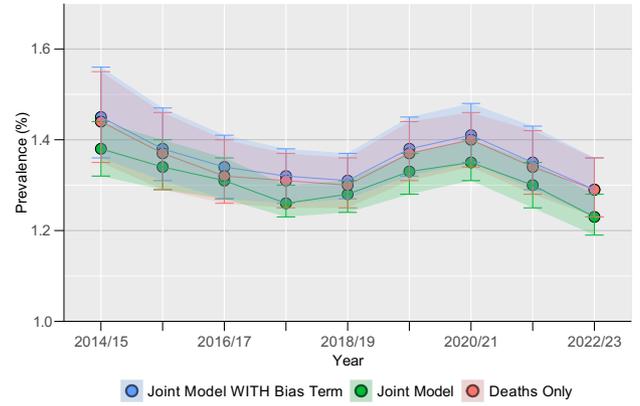

**Fig. 4.** Estimated prevalence (%) of opioid dependence in Scotland; Data Sources: Joint Model (both data sources), Joint Model with bias term, Deaths Only; 2014/15 to 2022/23

### 4.3. Incorporating external evidence

The MPEP approach can also be extended to incorporate external evidence or additional data. In the Scotland example, when prevalence estimates derived from a single event type (e.g., deaths only or hospital admissions only) differ markedly between data sources, this may indicate bias in one of the sources toward individuals already known to services. For instance, a non-fatal overdose may be more readily identified as opioid-related if it involves someone with a history of OAT. If external evidence were available to inform the bias parameter discussed in the previous subsection, it could be introduced in place of a vague prior, resulting in bias-adjusted estimates.

MPEP can also be extended in other ways to address data limitations. For example, in a previous application of MPEP in England, Jones et al. [17] had concerns about linkage between events and the cohort as it relied solely on initials and dates of birth. The authors were concerned that events occurring among individuals in the cohort might not be successfully linked. To address this issue, they introduced an additional parameter representing the probability of correct matching — referred to as 'pmatch' — which was incorporated into the regression models. Since some external information was available about the likelihood of misclassification due to unsuccessful linkage, an informative prior was assigned to 'pmatch'.

Including 'pmatch' in the model ensures that estimated event rates are adjusted for imperfect linkage to the cohort. Without this offset term, event rates could be underestimated, leading to inflated prevalence estimates [17]. If we wish to incorporate this offset into our model, it would be implemented as shown below.

- For the Poisson parameter among those within the cohort

$$\lambda^c_{igyrs} \cdot t_{gyrs}$$

$$\downarrow$$

$$pmatch \cdot \lambda^c_{igyrs} \cdot t_{gyrs}$$

The inclusion of the extra term in the model means that estimated event rates are adjusted to account for imperfect linkage between data sources. Since 'pmatch' takes values below one, the proposed adjustment implies that the number of observed events linked to the cohort represents only a proportion of the true number of events in the baseline cohort. Setting 'pmatch' to one would correspond to perfect linkage, resulting in no adjustment to the event rates.

- For the Poisson parameter among those not in the cohort

$$\lambda^c_{igyr0} \cdot t^e_{gyr}$$
$$\downarrow$$
$$pmatch \cdot \lambda^c_{igyr0} \cdot t^e_{gyr} + (1 - pmatch)\left(\lambda^c_{igyr0} \cdot t_{gyr0} + \lambda^c_{igyr1} \cdot t_{gyr1}\right)$$

where the term $\lambda^c_{igyr0} \cdot t_{gyr0} + \lambda^c_{igyr1} \cdot t_{gyr1}$ represents the number of events that occurred among individuals in the cohort but were not successfully linked. These unlinked events are therefore treated by the model as additional events 'outside the' cohort ('extra'). However, incorporating the 'pmatch' term into the model adjusts for imperfect linkage and helps prevent underestimation of event rates. Specifically, two additional terms are added to the expectation of extra events to represent events that occurred among individuals in the cohort but were missed due to incomplete linkage — and thus mistakenly classified as extra events. This approach ensures that only a portion of the extra events is attributed to the unobserved population (and accounted for in the extra event rate), while the remainder is assumed to have occurred within the baseline cohort.

For a demonstration of how this adjustment impacts the estimated prevalence, see Jones, et al., 2020 [17].

## 5. Discussion

Estimates of the size of populations of interest are crucial for policymakers and health officials to design interventions and policies that better address the needs of these communities. Indirect methods, such as the multiplier method and capture-recapture, are often employed to produce these estimates, using rich linked data from multiple sources [6]. However, these methods may have significant limitations arising from their underlying assumptions — for example, capture-recapture assumes no k-way interaction between data sources, with k being the number of data sources — assumptions that are often violated in practice [17, 23].

Here, we discuss an evidence synthesis model [1] that uses linked data to estimate the size of populations of interest in a generic way. We used a case study estimating the prevalence of opioid dependence in Scotland [22, 23] to demonstrate some of the model's features, but several of the adjustments and assumptions can be readily modified to better suit the data sources available in other settings. The MPEP approach is a flexible Bayesian approach that is easily adaptable to local contexts. By building upon linkages between administrative datasets — which are becoming increasingly available worldwide — it can be used to estimate the size of other populations of interest, including people dependent on stimulants such as cocaine and methamphetamine.

In this work, we discussed how the model can be adjusted to assess potential inconsistencies across different event types and propose several approaches to address these, if needed [28, 26]. As different case studies require different assumptions, some of the key assumptions of the model can be easily adapted to better fit specific contexts. For instance, we assume that the event rate among those not observed in the baseline cohort is equal to the event rate among those observed during periods off treatment. If analysts believe this assumption does not hold — for example, if the event rates differ in some way — they could incorporate informative priors to guide this difference, potentially even at specific levels of stratification. This might apply in situations where a treatment effect could not be directly accounted for within the baseline cohort, but there is external evidence about the event rate ratio between those within and outside the cohort.

As MPEP fits regression models on the event rates of observed events (assuming Poisson distributions for the observed counts), it may be the case that some events are sparse or that certain levels of stratification exhibit overdispersion or an excessive number of zeros. Here, we discuss how changes to the assumed distributions of the observed counts can be made to accommodate these features.

Implementation of MPEP requires a baseline cohort that includes individuals from the population of interest, and at least two type of event that occurs within this community. While prevalence estimation based on a single event type is possible, it does not allow for internal consistency checks and may result in less robust estimates. The definition of the 'baseline cohort', as well as any relevant stratification (such as by treatment status; if applicable), must be tailored to the available data and the specific context of each study. For example, in the case of estimating the number of people dependent on cocaine or methamphetamine there are a range of treatment options emerging but no 'standard pharmacotherapy' as for opioid dependence [20, 21]. Characterising the observed population cohort 'known' to be in treatment will still be important but the proportion of people in treatment is likely to be lower than for opoid dependents and there maybe other differences in relation to time in treatment and treatment effect that will need to be taken into account.

The event type must be highly specific to the population of interest, such that it occurs only among this group, in order to allow for unbiased estimation of the population size. As more data sources are integrated into the model (e.g., additional event types or other forms of evidence), the estimates of population size become more robust and reliable. The model's flexibility allows for some misclassification of events across observation groups and can even account for this, given appropriate supporting evidence [17].

Our aim is to further develop and adapt the model to estimate the prevalence of other types of drug dependence or related characteristics. There is a growing demand for robust and reliable estimates of the number of people dependent on substances such as cocaine and methamphetamine, as well as MPES models of HCV and HIV that incorporate robust estimates of population prevalence.

## A. Appendices

### A.1. Zero-Inflated Poisson

A zero-inflated Poisson model is used when a dataset contains significantly more zero counts than would be expected under a standard Poisson distribution. Assuming $X \sim ZIP(\lambda, \pi)$, the likelihood takes the form:

$$Pr(X = x) = L(x|\lambda, \pi) = \begin{cases} \pi + (1-\pi)e^{-\lambda} & \text{if } x = 0 \\ (1-\pi)\frac{\lambda^x e^{-\lambda}}{x!} & \text{if } x > 0 \end{cases}$$

The zero-inflated parameter $\pi$, represents the probability of observing a zero.

## A.2. Negative Binomial & Zero-Inflated Negative Binomial

The Poisson model assumes that the variance of the number of events in a given time interval is equal to the mean rate of occurrences. However, when the variance exceeds the mean — a phenomenon known as *overdispersion* — the negative binomial model is often used instead [12]. The negative binomial model accommodates this extra variability, making it a more flexible alternative to the Poisson distribution for modelling count data with greater dispersion. Assuming $X \sim NB(\lambda, \theta)$, the likelihood takes the form:

$$Pr(X = x) = L(x|\lambda, \theta) = \binom{x + \theta - 1}{x} \left(\frac{\theta}{\theta + \lambda}\right)^\theta \left(\frac{\lambda}{\theta + \lambda}\right)^x$$

$$= \frac{\Gamma(x + \theta)}{\Gamma(\theta)x!} \left(\frac{\theta}{\theta + \lambda}\right)^\theta \left(\frac{\lambda}{\theta + \lambda}\right)^x$$

The dispersion parameter $\theta$ quantifies the extent to which the variance exceeds the mean [?]. Specifically, a larger value of $\theta$ (significantly greater than zero) suggests that overdispersion is not present. As $\theta$ increases, the variance approaches the mean, and the Negative Binomial distribution converges to a Poisson distribution — i.e., if $\theta \to \infty$, then $\frac{1}{\theta} \to 0$ and $Var[X] \to \mu$. Conversely, a smaller value of $\theta$ indicates that the variance substantially exceeds the mean, confirming overdispersion and justifying the use of the Negative Binomial model instead of the Poisson.

A Zero-Inflated Negative Binomial (ZINB) model is used for count data that exhibit both overdispersion and excess zeros. Assuming $X \sim ZINB(\lambda, \theta, \pi)$, the likelihood takes the form:

$$L(x|\lambda, \theta, \pi) = \begin{cases} \pi + (1-\pi)\left(\frac{\theta}{\theta+\lambda}\right)^\theta & \text{if } x = 0 \\ (1-\pi)\frac{\Gamma(x+\theta)}{\Gamma(\theta)x!}\left(\frac{\theta}{\theta+\lambda}\right)^\theta\left(\frac{\lambda}{\theta+\lambda}\right)^x & \text{if } x > 0 \end{cases}$$

## A.3. Model Comparison

### A.3.1. Residual Deviance

The posterior mean residual deviance is defined as the difference between the deviance of the fitted model and that of a saturated model [36]. Formally,

$$ResDev = 2 \cdot (l(saturated) - l(fitted)),$$

where $l(saturated)$ is the log-likelihood of the saturated model, and $l(fitted)$ is the log-likelihood of the fitted model.

The Residual Deviance is a useful measure for assessing absolute model fit, helping to understand whether a model explain well the data. A well fitting model has posterior mean residual deviance approximately equal to the number of data points [36]. If the residual deviance is significantly larger than the number of data points, this suggests a poor model fit. In such cases, examining individual contributions to residual deviance can help identify data points that disproportionately influence the residual deviance, potentially indicating outliers or model misfit.

The formulas for the models discussed in this study are reported below.

- If $X_i \sim Poisson(\lambda_i)$, $i = 1, 2, ..., N$, then

$$ResDev = 2 \sum_{i=1}^{N} \left( \mathbf{1}_{x_i=0} \times \lambda_i + \mathbf{1}_{x_i>0} \times \left( x_i \log \frac{x_i}{\lambda_i} + (\lambda_i - x_i) \right) \right)$$

- If $X_i \sim NB(\lambda_i, \theta)$, $i = 1, 2, ..., N$, then

$$ResDev = 2 \sum_{i=1}^{N} \left( \mathbf{1}_{x_i=0} \times \left( \theta_i \log \frac{\theta_i + \lambda_i}{\theta_i} \right) + \mathbf{1}_{x_i>0} \times \left( \theta_i \log \frac{\theta_i + \lambda_i}{\theta_i + x_i} + x_i \log \left( \frac{x_i}{\lambda_i} \cdot \frac{\theta_i + \lambda_i}{\theta_i + x_i} \right) \right) \right)$$

- If $X_i \sim ZIP(\lambda_i, \pi)$, $i = 1, 2, ..., N$, then

$$ResDev = 2 \sum_{i=1}^{N} \left( \mathbf{1}_{x_i=0} \times \left( -\log\left(\pi + (1-\pi)e^{-\lambda_i}\right) \right) + \mathbf{1}_{x_i>0} \times \left( x_i \log \frac{x_i}{\lambda_i} + (\lambda_i - x_i) \right) \right)$$

- If $X_i \sim ZINB(\lambda_i, \theta, \pi)$, $i = 1, 2, ..., N$, then

$$ResDev = 2 \sum_{i=1}^{N} \left( \mathbf{1}_{x_i=0} \times \left( \log\left(\pi + (1-\pi)\left(\frac{\theta}{\theta + \lambda_i}\right)^\theta\right) \right) + \mathbf{1}_{x_i>0} \times \left( \theta_i \log \frac{\theta_i + \lambda_i}{\theta_i + x_i} + x_i \log \left( \frac{x_i}{\lambda_i} \cdot \frac{\theta_i + \lambda_i}{\theta_i + x_i} \right) \right) \right)$$

A useful metric for determining whether to include a term in a regression model is the combination of the Residual Deviance and the effective number of parameters (pD; see Appendix A.3.2). Models with a difference in the sum of these two components of less than three points are generally considered to have a comparable fit to the data [2, 36]. In this work, our model selection strategy involved sequentially adding potential interaction terms and retaining each only if its inclusion reduced the total measure by at least three points.

### A.3.2. Deviance Information Criterion

The Deviance Information Criterion (DIC) is a model selection metric that provides a balance between goodness of fit and model parsimony, assessing how well a model proposed predicts new data [9, 31].

The effective number of parameters of a model is estimated using the model's likelihood as [9, 31]:

$$pD = 2\left(\overline{l(y|\theta)} - l\left(y|\overline{\theta}\right)\right),$$

where $\overline{l(y|\theta)}$ is the posterior mean of the total log likelihood, and $l\left(y|\overline{\theta}\right)$ is the log likelihood evaluated at the posterior mean of the parameter $\theta$. The pD essentially represents the difference between the posterior mean deviance and the deviance of the posterior mean.

Then, the DIC could be defined as [9, 31]:

$$DIC = -2l\left(y|\overline{\theta}\right) + 2pD$$

$$= -2l(y|\theta) + pD$$

The DIC can be used to compare non-nested models by calculating the DIC for each model and selecting the model with the lowest DIC value, however, its interpretation is less clear than for nested models (e.g. it does not reflects the improvements of the model's fit as a result of the addiction of a covariate), and differences in DIC values might not be as informative.

## 6. Competing interests

No competing interest is declared.

## 7. Acknowledgments

This work was supported by funding from the Scottish Government, and from TRANSFORM (Transforming the evidence base on estimating prevalence of opioid use disorder and expanded access to interventions to prevent drug related deaths: an international data linkage study) NIH (1R01DA059822-01A1).